\documentclass[runningheads]{llncs}
\usepackage{epic,latexsym,amssymb}
\usepackage{amsfonts}
\usepackage{amscd}
\usepackage{amsmath}
\usepackage{graphicx}
\usepackage{color}

\usepackage{float}

\newtheorem{thm}{Theorem}

\newtheorem{ob}[thm]{Observation}
\newtheorem{prop}[thm]{Proposition}
\newtheorem{lem}[thm]{Lemma}
\newtheorem{quest}{Question}
\newtheorem{defn}{Definition}
\newtheorem{cor}[thm]{Corollary}

\usepackage[algo2e,ruled,vlined]{algorithm2e}

\begin{document}

\pagestyle{headings}
\mainmatter

\title{Characterizations of the Connected Forcing\\ Number of a Graph}
\titlerunning{On the Connected Forcing Number of a Graph}

\author{Boris Brimkov\inst{1} and Randy Davila\inst{2,3}}
\authorrunning{B. Brimkov and R. Davila}

\institute{
Department of Computational \& Applied Mathematics, Rice University, Houston, TX 77005, USA\\
\email{boris.brimkov@rice.edu}
\and
Department of Mathematics, Texas State University-San Marcos, San Marcos, TX 78666, USA\\
\email{rrd32@txstate.edu}
\and
Department of Mathematics, University of Johannesburg, Auckland Park 2006, South Africa\\
}


\maketitle

\begin{abstract}
Zero forcing is a dynamic graph coloring process whereby a colored vertex with a single uncolored neighbor forces that neighbor to be colored. This forcing process has been used to approximate certain linear algebraic parameters, as well as to model the spread of diseases and information in social networks. In this paper, we introduce and study the connected forcing process -- a restriction of zero forcing in which the initially colored set of vertices induces a connected subgraph. We show that the connected forcing number -- the cardinality of the smallest initially colored vertex set which forces the entire graph to be colored -- is a sharp upper bound to the maximum nullity, path cover number, and leaf number of the graph. We also give closed formulas and bounds for the connected forcing numbers of several families of graphs including trees, hypercubes, and flower snarks, and characterize graphs with extremal connected forcing numbers.

\smallskip


\end{abstract}

\section{Introduction}

In contrast to the static vertex and edge colorings classically studied in graph theory, recent years have seen the development of dynamic graph colorings, which are allowed to change in time according to predefined rules. One of the predominant dynamic colorings is the result of the \emph{forcing process}, defined as follows: given a simple graph $G=(V,E)$, let $R \subset V$ be an initial set of \emph{colored} vertices, all remaining vertices being \emph{uncolored}. At each integer valued time step, a colored vertex with a single uncolored neighbor will \emph{force} that neighbor to become colored; such a vertex is called a \emph{forcing vertex}. A set $R \subset V$ of initially colored vertices is called a \emph{forcing set} if, by iteratively applying the forcing process, all of $V$ becomes colored. The \emph{forcing number} of a graph $G$, denoted $F(G)$, is the cardinality of a smallest forcing set. 

Forcing on graphs was originally introduced in a workshop on linear algebra and graph theory in 2006 \cite{AIM-Workshop} and was used to bound the minimum rank of a graph.\footnote{Forcing sets were introduced by the name \emph{zero forcing sets}, and the forcing number was known as the \emph{zero forcing number}, denoted by $Z(G)$ in place of $F(G)$. We choose to conform to the newer nomenclature which appeared in a recent generalization of zero forcing to $k$-forcing (cf. \cite{k-Forcing}).} Namely, if $G$ is a graph whose vertices are labeled from $1$ to $n$, and $mr(G)$ denotes the minimum rank over all symmetric real valued matrices, where for $i\neq j$, the $ij^\text{th}$ entry is nonzero if and only if $(i,j)$ is an edge in $G$, then $mr(G)\geq n- F(G)$. Though the forcing number originated in relation to the minimum rank problem, forcing sets and the forcing number have since found a variety of applications in physics, logic circuits, coding theory, power network monitoring, and in modeling the spread of diseases and information in social networks; see \cite{quantum1,logic1,powerdom3,powerdom2}. The forcing number has also been used to bound or approximate various graph parameters \cite{zf_tw,zf_np}.

Computing $F(G)$ is $NP$-hard \cite{zf_np2}, though closed formulas and characterizations do exist for graphs with specified structure (see \cite{AIM-Workshop,benson,Eroh,Meyer}). One of the main obstacles in studying $F(G)$ is the fact that, in general, very little can be said about the structure of a forcing set. This difficulty motivated the investigations of the present paper. In particular, we inquire what can be said about the forcing process and forcing number of a graph, if the initial forcing set is assumed to be connected. More formally, if $R$ is a forcing set of $G$, and $R$ induces a connected subgraph, we say that $R$ is a \emph{connected forcing set}. The \emph{connected forcing number}, denoted $F_c(G)$, is the cardinality of a minimum connected forcing set in $G$, and is the main focus of this paper. In particular, we study the connected forcing number of several families of graphs including trees, flower snarks, hypercubes, and graphs with a single maximal clique of size greater than 2, with the hope of understanding the underlying structure of forcing sets in general, as well as exploring this newly defined graph invariant. Throughout this paper, we also recall various results related to forcing and other graph parameters, and compare them to our results about connected forcing.

We leave the computational complexity of connected forcing as an open question. Some of the examples and results given in this paper indicate that the connectivity of the forcing set can make the problem somewhat easier. On the other hand, we note that the related problem of graph domination and its connected variant are both NP-hard \cite{GJ}. However, it may be possible to efficiently compute the connected forcing number of families of graphs with polynomially many connected subgraphs, by adapting existing algorithms for computing the forcing number \cite{butler}.

This paper is organized as follows. In the next section, we recall some graph theoretic notions and notations. In Section 3, we show a number of relations between forcing and connected forcing in graphs, and prove several technical lemmas. In Section 4, we give closed forms and bounds for the connected forcing numbers of trees, graphs with a single maximal clique of size greater than~2, and flower snarks; we also characterize graphs with extreme connected forcing numbers. We conclude with some final remarks and open questions in Section~5.

\section{Preliminaries}

Let $G=(V,E)$ be a graph. The order and size of $G$ will be denoted by $n=|V(G)|$ and $m=|E(G)|$, respectively. Two vertices $v,w\in V$ are said to be adjacent, or neighbors, if there exists the edge $(v,w)\in E$. The open neighborhood of $v\in V$ is the set of all vertices which are adjacent to $v$, denoted  $N(v;G)$; the dependence on $G$ can be omitted when it is clear from the context. The degree of $v\in V$ is defined as $d(v)=|N(v)|$. The minimum degree and maximum degree of $G$ will be denoted as $\delta=\delta(G)$ and $\Delta=\Delta(G)$, respectively. 

A \emph{leaf} or \emph{pendant} is a vertex with degree 1. The \emph{leaf number} of $G$, denoted $L(G)$, is the number of leaves of $G$. The leaf number of a vertex $v$ is defined as
$L(v;G)=|\{\ell \in N(v;G): \ell \text{ is a leaf}\}|$, i.e., the number of leaves adjacent to $v$. A \emph{path cover} of $G$ is a set of vertex-disjoint induced paths in $G$ which contain all the vertices of $G$. The \emph{path cover number} of $G$, denoted $P(G)$, is the minimum size of a path cover. The \emph{maximum nullity} of $G$ with vertex set $\{1,\ldots,n\}$ is defined as the maximum nullity over all symmetric real valued matrices, where for $i\neq j$, the $ij^\text{th}$ entry is nonzero if and only if $(i,j)$ is an edge in $G$, i.e., $M(G)=\max\{null(A):A\in\mathcal{S}(G)\}$.

A \emph{clique} of $G$ is a complete subgraph; for our purposes, it will be convenient to think of a clique as a maximal (with respect to inclusion) complete subgraph. An \emph{articulation point} (also called a \emph{cut vertex}) is a vertex which, when removed, increases the number of connected components in $G$. Similarly, a \emph{bridge} (also called a \emph{cut edge}) is an edge which, when removed, increases the number of components of $G$. A \emph{biconnected component} or \emph{block} of $G$ is a maximal subgraph of $G$ which has no articulation points. Because disconnected graphs can never have a connected forcing set, we will consider only connected graphs for the remainder of this paper. For other graph theoretic terminology and definitions, we refer the reader to \cite{bondy}. We are now ready to present our study of connected forcing.

\section{Relation of connected forcing to other parameters}

In this section, we study relations between the connected forcing number and other graph parameters such as the forcing number, maximum nullity, path cover number, and leaf number. We also compare several properties of minimum connected forcing sets and minimum forcing sets. In later sections, we give several additional relations between forcing and connected forcing.

Our first simple observation is that since any connected forcing set is also a forcing set, any minimum connected forcing set will contain a (not necessarily minimum) forcing set. Thus, we have the following result.

\begin{ob}
\label{geq_fact}
For any connected graph $G$, $F_c(G)\geq F(G)$, and this bound is sharp.
\end{ob}

Graphs for which Observation \ref{geq_fact} is sharp include paths, cycles, and complete graphs. The following observation can help us come up with other, non-trivial examples.

\begin{ob}
\label{iff_fact}
For any connected graph $G$, $F(G)=F_c(G)$ if and only if there exists some minimum forcing set of $G$ which is connected.
\end{ob}

Note that Observation \ref{iff_fact} is not a tautology, since a graph could have minimum forcing sets which are not connected. In general, it is harder to characterize or even count all distinct minimal forcing sets of a graph than to find its forcing number. Thus, it may be hard to determine whether or not a graph has a minimum forcing set which is connected.

We can apply Observation \ref{iff_fact} to some known results about forcing and obtain analogous results about connected forcing. For example, Theorem 3.1 of \cite{AIM-Workshop} states that the forcing number of the hypercube graph $Q_n$ is $2^{n-1}$. Since there exists a connected set of size $2^{n-1}$ which forces $Q_n$ -- namely a subgraph of $Q_n$ isomorphic to $Q_{n-1}$ -- we conclude that $F_c(Q_n)=F(Q_n)=2^{n-1}$. As another example, Theorem 2.4 of \cite{benson} states that the forcing number of the torus graph $C_n\times C_m$, $m\geq n$, is $2n$. Since there exists a connected set of size $2n$ which forces $C_n\times C_m$ -- namely a subgraph of $C_n\times C_m$ isomorphic to $C_n\times P_2$ -- we conclude that $F_c(C_n\times C_m)=F(C_n\times C_m)=2n$. The same approach can be attempted for other families of graphs whose forcing number has been characterized. Of course, as discussed above, this simple approach will not always work, as demonstrated by later examples in the paper.

Since the forcing number of a graph is an upper bound on the maximum nullity and path cover number of the graph (cf. \cite{AIM-Workshop,Hogben}), Observation \ref{geq_fact} also yields the following relations.

\begin{ob}
\label{M_obs}
For any connected graph $G$, $F_c(G)\geq M(G)$, and this bound is sharp.
\end{ob}

\begin{ob}
\label{P_obs}
For any connected graph $G$, $F_c(G)\geq P(G)$, and this bound is sharp.
\end{ob}

Observations \ref{M_obs} and \ref{P_obs} are sharp, e.g., for paths. We defer further study of the relations between connected forcing, maximum nullity, and path covers to future work. 
Our next result shows that the connected forcing number is also an upper bound on the number of leaves in the graph. Note that an analogous relation does not always hold for the forcing number (see, e.g., Figure 1, right).

\begin{prop}
\label{leaf_prop}
For any connected graph $G$ different from a path, $F_c(G)\geq L(G)$, and this bound is sharp.
\end{prop}

\proof
Let $R$ be an arbitrary connected forcing set of $G$. If a vertex forces another vertex at some step in the forcing process, then it cannot force a second vertex at a later time, since that would imply it had two uncolored neighbors when it forced for the first time. Thus, each sequence of forces induces a path in $G$. More precisely, for each vertex $v\in R$, there is an induced path $P$ with $v$ being one of the ends of $P$, and all other vertices of $P$ being uncolored at the initial time step. Such a path is called the \emph{forcing chain} corresponding to $v$. Each vertex, and in particular each leaf of $G$ has to be in some forcing chain. Of course, a chain cannot contain more than two leaves, since then it wouldn't induce a path. Suppose some chain contains two leaves. Then one of these leaves must be in $R$, and every other vertex in the chain must not be in $R$. In particular, the neighbor of the colored leaf is uncolored. However, since $G$ is not a path, there must be other members of $R$ outside of this forcing chain; thus, $R$ is not connected -- a contradiction. Thus, each forcing chain can contain at most one leaf. Since each forcing chain contains one element of $R$, it follows that $F_c(G)\geq L(G)$. This bound is sharp, e.g., for the graph in Figure \ref{fig2}, left.
\qed

In contrast to graphs for which $F_c(G)=F(G)$, there are also families of graphs for which the connected forcing number can be arbitrarily larger than the forcing number. We state this more precisely below. 

\begin{figure}[h!]
\begin{center}
\includegraphics[scale=0.5]{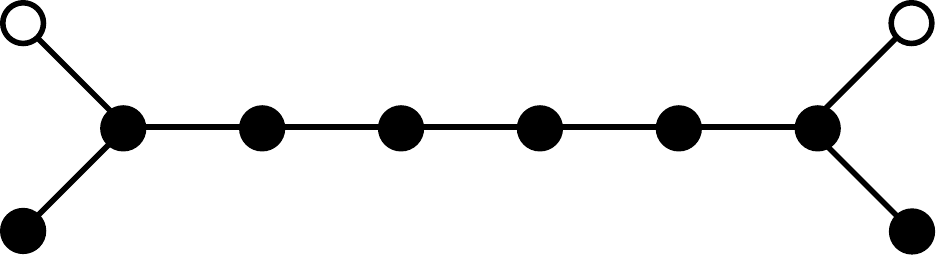}\qquad \qquad
\includegraphics[scale=0.5]{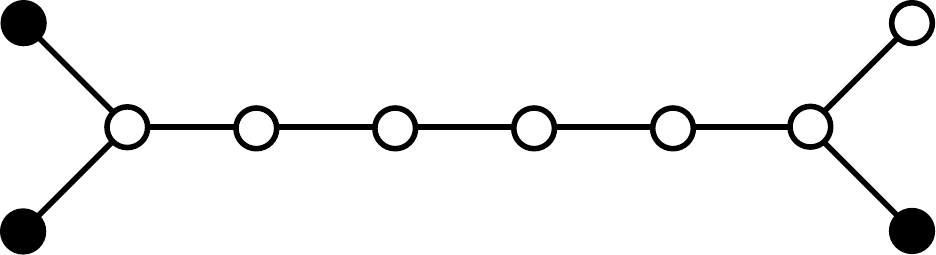}
\caption{\emph{Left:} Minimum connected forcing set. \emph{Right:} Minimum forcing set.}
\label{fig1}
\end{center}
\end{figure}

\begin{figure}[h!]
\begin{center}
\includegraphics[scale=0.55]{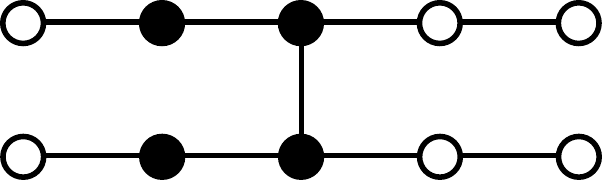}\qquad \qquad
\includegraphics[scale=0.55]{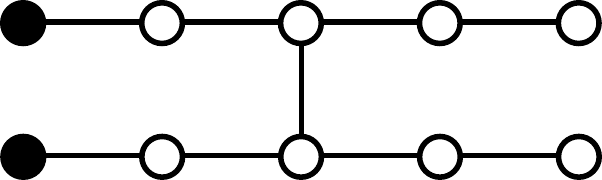}
\caption{\emph{Left:} Minimum connected forcing set. \emph{Right:} Minimum forcing set.}
\label{fig2}
\end{center}
\end{figure}

\begin{ob}
For any $a>0$, there exists a graph $G$ such that $F_c(G)\geq F(G)+a$.
\end{ob}
\proof
Consider the graph $G_k$ obtained by attaching a pair of pendant vertices to each end of a path $P_k$. The forcing number of $G_k$ is 3, and for $k\geq 2$ its connected forcing number is $k+2$; see Figure \ref{fig1} for an illustration.
\qed

Since any minimum connected forcing set contains a forcing set, a natural question to ask is whether every graph has some minimum connected forcing set that contains a minimum forcing set. Analogously, one could ask whether a minimum forcing set can be extended to obtain a minimum connected forcing set. However, the graph in Figure \ref{fig2} is a counterexample to both of these questions. In particular, no subset of any minimum connected forcing set of the graph is a minimum forcing set, and no superset of any minimum forcing set of the graph is a minimum connected forcing set. We state this formally in Observation \ref{ob_subset}.

\begin{ob}
\label{ob_subset}
A minimum connected forcing set of a connected graph $G$ does not necessarily contain a minimum forcing set of $G$.
\end{ob}

An important concept to studying and understanding the forcing process is that of the \emph{forcing spread} of a vertex $v$; this parameter, defined to be $f(G;v)=F(G)-F(G-v)$, describes the effects of deleting a vertex from the graph on the forcing number of the graph. It has been shown in \cite{Edholm,Huang} that the forcing spread of a vertex is bounded by 1; more precisely, $-1\leq f(G;v)\leq 1$. We define the analogous concept of \emph{connecting forcing spread} of a vertex $v$ as $f_c(G;v)=F_c(G)-F_c(G-v)$. In this definition, we restrict $v$ to be a non-articulation point of $G$, since the connected forcing number is undefined for a disconnected graph. With this in mind, we show that unlike the forcing spread, the connected forcing spread of a vertex can be arbitrarily large. 

\begin{ob}
For any $c_1<0$ and $c_2>0$, there exist graphs $G_1$ and $G_2$ and vertices $v_1\in G_1$ and $v_2\in G_2$ such that $f_c(G_1;v_1)<c_1$ and $f_c(G_2;v_2)>c_2$. 
\end{ob}

\begin{figure}[h!]
\begin{center}
\includegraphics[scale=0.5]{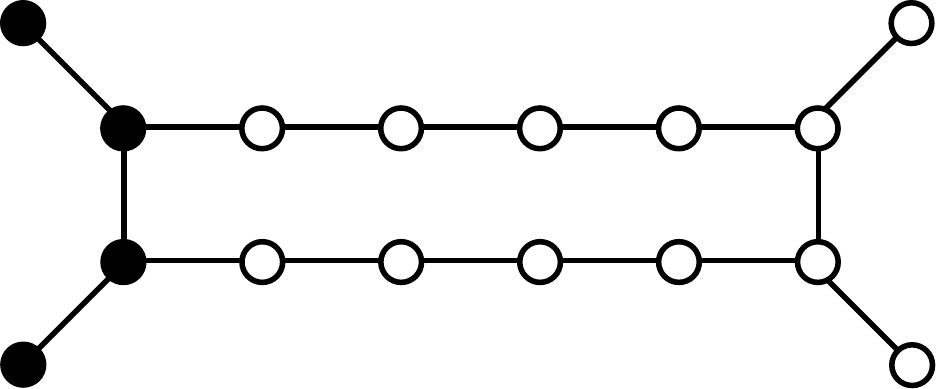}\qquad \qquad
\includegraphics[scale=0.5]{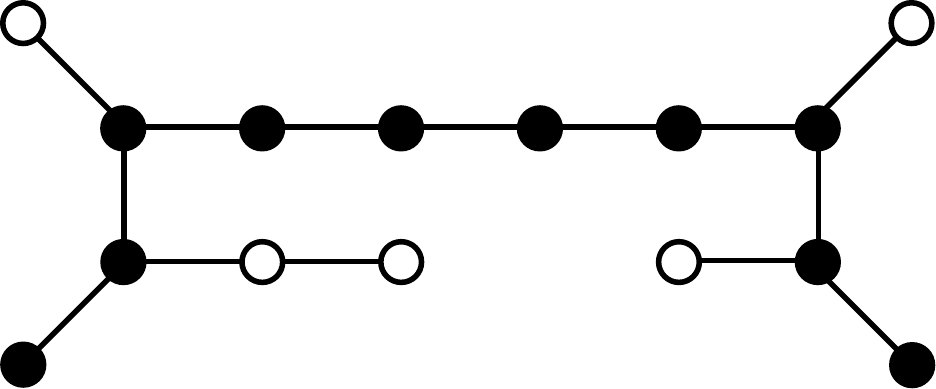}
\caption{Deleting a vertex can make the connected forcing number grow arbitrarily.}
\label{fig3}
\end{center}
\end{figure}

\proof
Consider the graph $H_k$ obtained by appending a pendant vertex to each endpoint of two maximally distant edges of an even cycle $C_k$. It is easy to verify that $F_c(H_k)=4$. Now let $v$ be a vertex at distance at least 2 from a pendant. It is easy to verify that $F_c(H_k-v)=\frac{k}{2}+4$ for $k\geq 10$. Thus, $f_c(H_k;v)=-\frac{k}{2}$; see Figure \ref{fig3} for an illustration. 

Let $G_k$ be the family of graphs described in Observation 5 and pictured in Figure \ref{fig1}. It was shown earlier that $F_c(G_k)=k+2$. Let $v$ be a leaf of $G_k$. It is easy to see that $F_c(G_k-v)=3$. Thus, $f_c(G_k;v)=k-1$.
\qed

In studying the forcing process, it is important is to consider what a single vertex can force. It is easy to see that a connected forcing set of size 1 can force a graph if and only if the graph is a path. The following lemma gives conditions under which only a path subgraph can be forced.

\begin{lem}
\label{technical_lemma1}
Let $G=(V,E)$ be a connected graph, $v$ be an articulation point, $V_1$ be the vertex set of a connected component of $G-v$, and $V_2=V-V_1-\{v\}$. Suppose $v$ and the vertices in $N(v)\cap V_1$ are colored, and every vertex in $V_2$ is uncolored. Then $G$ can be forced only if $G[V_2]$ is a path.
\end{lem}

\proof
Suppose $G[V_2]$ is not a path. If $|N(v)\cap V_2|\geq 2$, $v$ cannot force any vertex in $V_2$ since it has at least two uncolored neighbors; also, no vertex in $V_1$ can force a vertex in $V_2$ since any forcing chain to $V_2$ must pass through $v$. Thus, in this case, $G$ cannot be forced. If $|N(v)\cap V_2|=1$, let $P=\{v,p_1,\ldots,p_k,w\}$ be a maximal set of vertices in $V_2\cup\{v\}$ such that $G[P]$ is a path with one endpoint at $v$ and the other endpoint at $w$, and such that $d(p_i)=2$, $1\leq i\leq k$. Then $d(w)\geq 3$, since if $d(w)=1$, $G[V_2]$ would be a path, and if $d(w)=2$, $P$ would not be maximal. Then $v$ can force all vertices in $P$ up to $w$, but $w$ cannot force any vertex in $V_2-P$, since $w$ has at least two uncolored neighbors. Also, no vertex in $V_1\cup P$ can force a vertex in $V_2-P$ since any forcing chain to $V_2-P$ must pass through $w$. Thus, $G$ cannot be forced if $G[V_2]$ is not a path.
\qed

Our next result concerns the effect of subdividing an edge on the connected forcing number. In general, subdivision can increase the connected forcing number: consider, for example, the graph in Figure 1; subdividing any edge incident to a vertex of degree 2 will clearly increase $F_c(G)$. Subdivision can also increase the forcing number: consider, for example, the graph in Figure \ref{fig2}; subdividing the edge whose endpoints have degree 3 will increase $F(G)$. However, we prove below that subdividing a leaf edge does not affect the connected forcing number of the graph. We also prove that a leaf vertex incident to a vertex of degree 2 does not belong to any minimum connected forcing set.

\begin{lem}
\label{leaf_lemma}
Let $G=(V,E)$ be a connected graph different from a path and $\ell$ be a leaf adjacent to a vertex of degree 2. Then $\ell$ does not belong to any minimum connected forcing set of $G$, and $F_c(G)=F_c(G-\ell)$.
\end{lem}

\proof
Let $R$ be an arbitrary minimum connected forcing set of $G$. Let $v$ be the neighbor of $\ell$, and $w$ be the other neighbor of $v$. We will first show that $\ell\notin R$. Suppose on the contrary that $\ell\in R$, $v\in R$, and $w\in R$; then $R-\{\ell\}$ is a connected forcing set of smaller size, since $v$ can force $\ell$ in the first step of the forcing process. Next, suppose that $\ell\in R$, $v\in R$, and $w\notin R$. Since $G-\ell-v$ is not a path, by Lemma \ref{technical_lemma1}, some vertex in $V-\{\ell,v\}$ must be in $R$. But since $w\notin R$, $R$ will be disconnected -- a contradiction. The case when $\ell\in R$, $v\notin R$, and $w\notin R$ is handled analogously, and the case $\ell\in R$, $v\notin R$, $w\in R$ immediately contradicts $R$ being connected. Thus, $\ell$ is not in $R$. 

Now, if $R$ is a minimum connected forcing set of $G$, then $R$ is also a minimum connected forcing set of $G-\ell$, since $\ell$ cannot be initially colored and cannot be a non-terminal vertex of any forcing chain. Conversely, if $R$ is a minimum connected forcing set of $G-\ell$ and $v\in R$, then $w$ must also be in $R$ since otherwise $R$ will be disconnected; then, $v$ will force $\ell$ in $G$ in the first step of the forcing process. If $v\notin R$, then at some step of the forcing process $w$ will force $v$, and the addition of $\ell$ can have no effect on the forcing path of $w$ or any other vertex in $G$; at the next step of the forcing process, $v$ will force $\ell$. Thus, in both cases, $R$ is also a minimum connected forcing set of $G$, so $F_c(G)=F_c(G-\ell)$.\qed

We now fix some notation which will be used in the sequel. 

\begin{defn}
Let $G=(V,E)$ be a connected graph. Define 
\begin{eqnarray*}
R_1(G)&=&\{v\in V: G-v \text{ has at least 3 connected components}\}\\
R_2(G)&=&\{v\in V: G-v \text{ has 2 connected components, neither of which is a path}\}\\
R_3(G)&=&\{v\in V: v \text{ is adjacent to at least one leaf}\}.
\end{eqnarray*} 
When there is no scope for confusion, the dependence on $G$ will be omitted.
\end{defn}

\begin{defn}
Let $G$ be a connected graph different from a path. Then, $\bar{G}$ is the graph obtained by repeatedly deleting a leaf of $G$ which is adjacent to a vertex of degree 2, until no more such leaves can be deleted. 
\end{defn}

By inductively applying Lemma \ref{leaf_lemma}, we conclude that $F_c(G)=F_c(\bar{G})$ for any graph $G$. Moreover, it is easy to see that $R_1(G)=R_1(\bar{G})$, $R_2(G)=R_2(\bar{G})$, and $L(G)=L(\bar{G})$.

We now present another result related to leaves which must necessarily belong to a connected forcing set.

\begin{lem}
\label{leaf_lemma2}
Let $G=(V,E)$ be a connected graph different from a path and $R$ be an arbitrary connected forcing set of $G$. $R$ must contain all-but-one leaves adjacent to $v$ for each $v\in R_3(\bar{G})$.
\end{lem}

\proof
By Lemma \ref{leaf_lemma}, $R$ is also a minimum connected forcing set of $\bar{G}$. Suppose that for some $v\in R_3(\bar{G})$, there are two leaves $u$ and $w$ adjacent to $v$ that are not in $R$. Since $G$ is not a path, there must be another biconnected component incident to $v$. If one or both of $v$ and $w$ are in $R$ but  $v$ is not in $R$, we get a contradiction by Lemma \ref{technical_lemma1}, since then $R$ must either not be connected or not be forcing. If neither $u$ nor $w$ are in $R$, then they must be forced by $v$. But then we again get a contradiction by Lemma \ref{technical_lemma1}, since $v$ can only force a single leaf. Thus, for each $v\in R_3(\bar{G})$, all-but-one leaves adjacent to $v$ must be in $R$.  
\qed

Considering Lemma \ref{leaf_lemma2}, the next definition specifies the total number of leaves of $\bar{G}$ which must be in every minimum connected forcing set of $G$.

\begin{defn}
\label{defL}
Let $G=(V,E)$ be a connected graph. Then, $\mathcal{L}(G)=\sum_{v\in R_3(\bar{G})}(L(v;\bar{G})-1)$; when there is scope for confusion, the dependence on $G$ will be omitted.
\end{defn}
Our last result in this section again concerns the composition of connected forcing sets. It is shown in \cite{Barioli} that for any non-trivial connected graph $G$, no vertex belongs to every minimum forcing set. In contrast, we show that certain vertices do belong to every minimum connected forcing set of $G$.

\begin{lem}
\label{technical_lemma}
Let $G=(V,E)$ be a connected graph. If $v\in R_1$ or $v\in R_2$, then $v$ is contained in every minimal connected forcing set of $G$.
\end{lem}

\proof
Let $R\subset V$ be an arbitrary connected forcing set of $G$ and suppose that $v\in R_1$ or $v\in R_2$ but $v\notin R$. Suppose, for contradiction, that one connected component of $G-v$, say $G_1$, contains all the vertices of $R$; let $G_2=G[V-V(G_1)-\{v\}]$ be the other component(s) of $G-v$. By definition $v$ is an articulation point, so every forcing chain from a vertex of $G_1$ to a vertex of $G_2$ passes through $v$. Since $R$ is a forcing set of $G$, $v$ and every neighbor of $v$ in $G_1$ will become colored at some step in the forcing process. At this step, no vertex in $G_2$ can be colored, since $v$ cannot begin to force any vertices before all its neighbors in $G_1$ are colored. In the next steps of the forcing process, by Lemma \ref{technical_lemma1}, $v$ can only force a single path in $G_2$. Since $v\in R_1$ or $v\in R_2$, $G_2$ is not a single path, so the entire graph cannot be forced by $R$. Thus, at least two connected components of $G-v$ must contain vertices of $R$. But in this case, since $v\notin R$, $R$ cannot be connected. Thus, $v$ must be in $R$. \qed

We leave as an open problem determining the validity of the converse of Lemma \ref{technical_lemma}; in other words, is there a graph $G$ and a vertex $v$ contained in every minimal connected forcing set of $G$, for which $G-v$ has 2 connected components, one of which is a path?

\section{Characterizations of connected forcing numbers}

\subsection{Connected forcing in a tree}
In this section, we present several results on the connected forcing of trees. We begin with a closed form for $F_c(T)$. This result is constructive, and can be used to find a minimum connected forcing set in linear time. 
 
 

\begin{thm}
\label{tree_thm}
Let $T=(V,E)$ be a tree. Then,
\begin{equation*}
F_c(T)=\begin{cases}
      1 & \text{if }\Delta(T)<3 \\
      |R_1|+|R_2|+\mathcal{L} & \text{if }\Delta(T)\geq 3.
    \end{cases}
\end{equation*}
The following Procedure gives a minimum connected forcing set of $T$ in $O(n)$ time.
\begin{enumerate}
\item If $R_1=\emptyset$, color one vertex with degree $\delta$;
\item Color all vertices in $R_1$, $R_2$;
\item For each $v\in R_3(\bar{T})$, color all-but-one leaves adjacent to $v$.
\end{enumerate}
\end{thm}

\proof
If $\Delta(T)<3$, then $T$ is either a path or a single vertex and $F_c(T)=1$; in both cases, coloring one vertex with degree $\delta$ gives a minimum connected forcing set. Thus, assume henceforth that $\Delta(T)\geq 3$. 

Let $S\subset V$ be a set of vertices colored by the Procedure. We will show that $S$ is a minimum connected forcing set for $\bar{T}$, and by Lemma \ref{leaf_lemma}, this will imply that $S$ is a minimum connected forcing set for $T$.
All vertices of $\bar{T}$ with degree at least 3 are in $R_1$, and by construction of $\bar{T}$, all vertices of $\bar{T}$ with degree 2 are in $R_2$. Thus, the only vertices of $\bar{T}$ which are not initially colored are leaves, so $S$ is a connected set. Moreover, each uncolored leaf is adjacent to some vertex $v\in R_3(\bar{T})$, and since $R_3\subset R_1$, $v$ is colored, along with all-but-one of its neighbors. Thus, $S$ is a forcing set of $\bar{T}$. 

Let $R$ be an arbitrary minimum forcing set of $\bar{T}$. By Lemma \ref{technical_lemma}, all vertices in $R_1$ and $R_2$ must be in $R$, and by Lemma \ref{leaf_lemma2}, all-but-one leaves adjacent to each $v\in R_3(\bar{T})$ must be in $R$. Thus, $R$ must include all vertices colored by the Procedure, so $S$ is indeed a minimum connected forcing set. 

Since $R_1$, $R_2$, and the set of leaves of $\bar{T}$ are disjoint sets, and by Definition \ref{defL}, the number of leaves colored is $\mathcal{L}$, the number of vertices colored by the Procedure, and hence $F_c(T)$, is  $|R_1|+|R_2|+\mathcal{L}$. 

\qed

We now give a formula to count the number of distinct minimum connected forcing sets of a tree.

\begin{cor}
If $T\neq P_n$, there are $\prod_{v\in R_3(\bar{T})} L(v;\bar{T})$ distinct minimum connected forcing sets of $T$.
\end{cor}

\proof
In Theorem \ref{tree_thm} we have shown that every minimum connected forcing set of $T$ must contain all vertices in $R_1$ and $R_2$, and all-but-one leaves adjacent to $v$ for each $v\in R_3(\bar{T})$. For each $v\in R_3(\bar{T})$, there are $\binom{L(v;\bar{T})}{L(v;\bar{T})-1}=L(v;\bar{T})$ choices of which vertices to include in a minimum connected forcing set. Since these choices can be made independently for each $v\in R_3(\bar{T})$, there are $\prod_{v\in R_3(\bar{T})} L(v;\bar{T})$ distinct minimum connected forcing sets of $T$.
\qed

Our next result is a characterization of trees for which $F_c(T)=F(T)$. 
\begin{prop}
\label{tree_iff_prop}
Let $T$ be a tree. Then, $F_c(T)=F(T)$ if and only if $T=P_n$. 
\end{prop}

\proof
If $T=P_n$, it is easy to see that $F_c(T)=1=F(T)$. If $T\neq P_n$, then $F(T)\leq L(T)-1\leq F_c(T)-1<F_c(T)$, where the first inequality follows from Theorem 5.6 in \cite{k-Forcing} and the second inequality follows from Proposition \ref{leaf_prop}. By contraposition, $F_c(T)=F(T)$ implies $T=P_n$.
\qed

We conclude with a characterization of the minimum forcing sets of trees, which follows from Proposition \ref{tree_iff_prop} and Observation \ref{iff_fact}.

\begin{cor}
Every minimum forcing set of a tree $T\neq P_n$ is disconnected. 
\end{cor}

\subsection{Connected forcing in a graph with a single clique of size greater than 2}

We now give a closed form for the connected forcing number of graphs with a single maximal clique of size greater than 2. This result is constructive, and can be used to find a minimum connected forcing set in linear time.

\begin{thm}
Let $G=(V,E)$ be a connected graph with a single maximal clique of size greater than 2, which has vertex set $K$. Then,


\begin{equation*}
F_c(G)=\begin{cases}

      |R_1\cup R_2\cup K|+\mathcal{L}-1 & \text{if }K-R_3(\bar{G})\neq \emptyset \text{ and }K-R_1-R_2\neq \emptyset\\
      |R_1\cup R_2\cup K|+\mathcal{L }& \text{if }K-R_3(\bar{G})= \emptyset \text{ or }K-R_1-R_2= \emptyset
    \end{cases}
\end{equation*}

The following Procedure gives a minimum connected forcing set of $G$ in $O(n)$ time.
\begin{enumerate}
\item Color all vertices in $R_1$ and $R_2$;
\item If $K-R_3(\bar{G})\neq \emptyset$ and $K-R_1-R_2\neq \emptyset$, pick $w\in K-R_1-R_2$ and color all vertices in $K$ except $w$. Otherwise, color all vertices in $K$;
\item For each $v\in R_3(\bar{G})$, color all-but-one leaves adjacent to $v$.
\end{enumerate}
\end{thm}

\proof

Let $S$ be a set of vertices colored by the Procedure. We will show that $S$ is a minimum connected forcing set for $\bar{G}=(\bar{V},\bar{E})$ and by Lemma \ref{leaf_lemma}, this will imply that $S$ is a minimum connected forcing set for $G$.

All vertices of $\bar{V}-K$ are endpoints of cut edges, so they are either in $R_1$, or in $R_2$, or leaves. Thus, the only vertices in $\bar{V}-K$ that can be in $\bar{V}-S$ are leaves. The only vertex in $K$ that can be in $\bar{V}-S$ is a vertex $w\in K-R_1-R_2$, i.e., a non-articulation point of $K$, or an articulation point of $K$ incident to a single leaf, along with that leaf. Deleting all possible vertices in $\bar{V}-S$ from $\bar{G}$ does not disconnect $\bar{G}$, so $S$ is connected. 

If every vertex of $K$ is incident to a leaf, i.e., $K-R_3(\bar{G})=\emptyset$, or if every vertex of $K$ is in $R_1$ or $R_2$, i.e., $K-R_1-R_2=\emptyset$, then coloring all vertices in $K$ and all-but-one leaves adjacent to each $v\in R_3(\bar{G})$ will force all uncolored leaves adjacent to $K$. Likewise, coloring all vertices in $R_1$ and $R_2$ and all-but-one leaves adjacent to each $v\in R_3(\bar{G})$ will force all uncolored leaves in $\bar{V}-K$. Thus, in this case, $S$ is a forcing set.
 
On the other hand, if there is a vertex $v$ in $K$ which is not incident to a leaf, i.e., $v\in K-R_3(\bar{G})$, and a vertex $w$ in $K$ which is not an articulation point or is an articulation point incident only to $K$ and a single leaf, i.e., $w\in K-R_1-R_2$, then coloring every vertex in $K$ except $w$ will make $v$ able to force $w$. Then, by the same argument as in the previous case, it follows that $S$ is a forcing set.

Now let $R$ be an arbitrary minimum forcing set of $\bar{G}$. By Lemma \ref{technical_lemma}, all vertices in $R_1$ and $R_2$ must be in $R$. Suppose that two vertices $u$ and $v$ in $K$ are not in $R$. Then, each colored vertex in $K$ will have at least two uncolored neighbors, so no vertex in $K$ can force $u$ and $v$. Moreover, no vertex outside of $K$ can force $u$ and $v$, since otherwise, by Lemma \ref{technical_lemma1}, $R$ will not be connected. Thus, $R$ can exclude at most one vertex of $K$.

If all vertices of $K$ are adjacent to leaves, and one vertex $v$ in $K$ is not in $R$, then every colored vertex in $K$ will have two uncolored neighbors, so no vertex in $K$ will be able to force another vertex, and no vertex outside of $K$ will be able to force a leaf incident to $K$ or the uncolored vertex $v$. Thus, in this case, all vertices of $K$ must be in $R$. Alternately, if not all vertices of $K$ are adjacent to leaves but all vertices of $K$ are in $R_1$ or $R_2$, then again all vertices of $K$ must be colored by Lemma \ref{technical_lemma}. Finally, by Lemma \ref{leaf_lemma2}, for each $v\in R_3(\bar{G})$, all-but-one leaves adjacent to $v$ must be in $R$. Since $R$ must include all vertices colored by the Procedure, $S$ is indeed a minimum connected forcing set. 

If $K-R_3(\bar{G})\neq \emptyset$ and $K-R_1-R_2\neq \emptyset$, then $R$ contains all-but-one vertices of $K$ and all vertices in $R_1$ and $R_2$, i.e., $|R_1\cup R_2\cup K|-1$ vertices, as well as all-but-one leaves adjacent to each $v\in R_3(\bar{G})$, i.e., $\mathcal{L}$ leaves. If $K-R_3(\bar{G})=\emptyset$ or $K-R_1-R_2=\emptyset$, then $R$ contains all vertices of $K$, $R_1$, and $R_2$, i.e., $|R_1\cup R_2\cup K|$ vertices and $\mathcal{L}$ leaves. Thus, the connected forcing number of $G$ is as claimed. 
\qed

\subsection{Flower Snarks}

\emph{Snarks} are connected bridgeless cubic graphs with edge chromatic number $4$, and have been related to many graph theoretic properties. In this section, we give an upper bound for the connected forcing number of a special class of snarks called \emph{flower snarks}. Our investigation of snarks was prompted by the observation that the Petersen graph, which is a snark, has a minimum forcing set which is connected. We conjecture that this holds true for other snarks, as well.

A flower snark on $n$ vertices, denoted $J_n$, is constructed by the following process: 
\begin{enumerate}
\item Start with $k$ (odd) copies of the star $K_{1,3}$. Denote the central vertex of each star $A_j$, and the outer vertices by $B_j$, $C_j$, and $D_j$. 

\item Construct the $k$-cycle $(B_1,...,B_k)$.

\item Finally, construct the $2k$-cycle $(C_1,...,C_k,D_1,...,D_k)$.
\end{enumerate}

\begin{prop}\label{snark upper}
Let $J_n$ be a flower snark with order $n=4k$. Then, $F_c(J_n)\le \frac{n}{4} + 2$.

\end{prop}

\begin{figure}[h!]
\begin{center}
\includegraphics[scale=0.17]{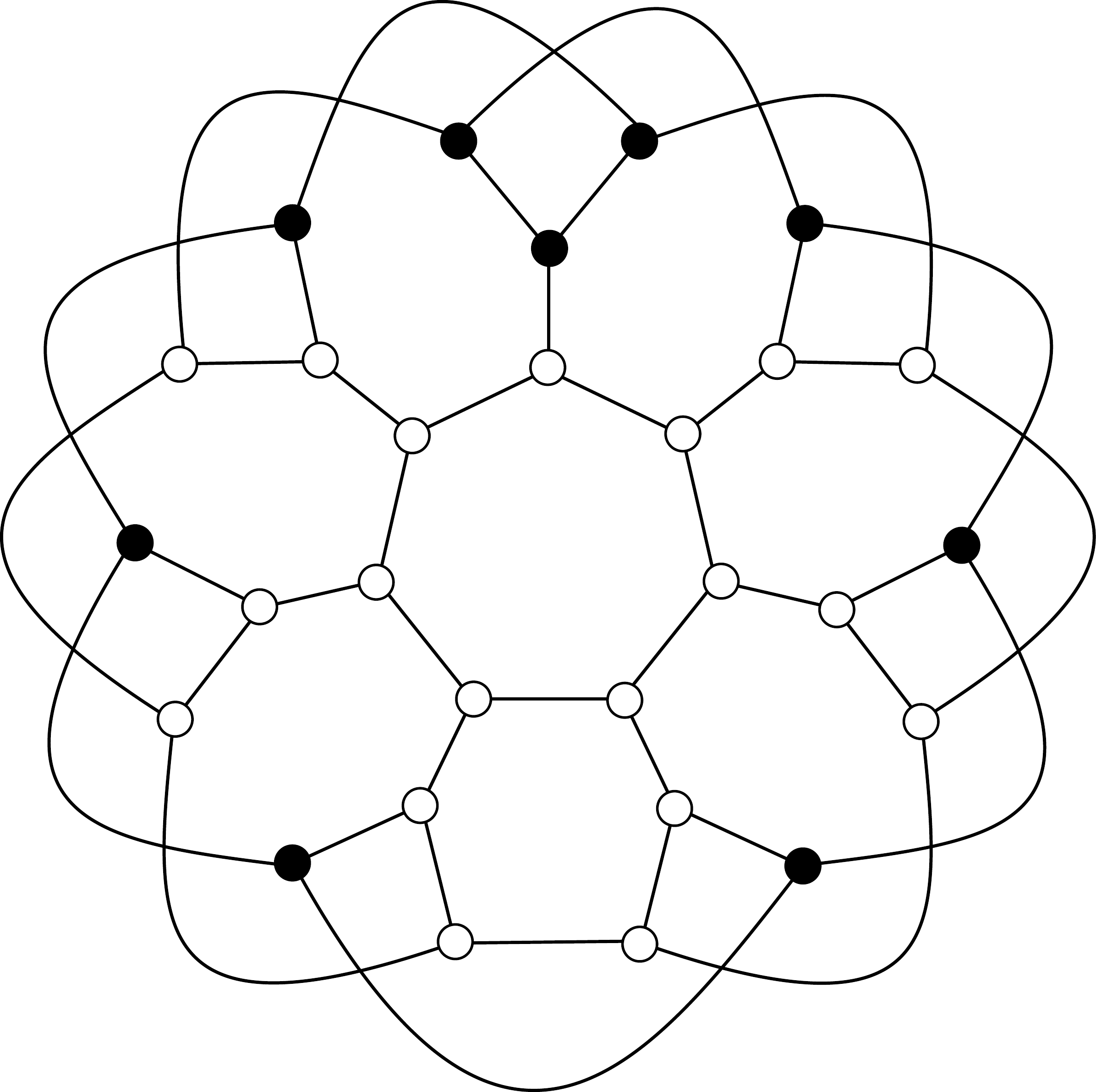}
\caption{ A connected forcing set of the flower snark $J_{28}$.}
\label{fig4}
\end{center}
\end{figure}

\proof Let $J_n$ be a flower snark on $n=4k$ vertices. Color $C_1,...,C_k$, $A_1$, and $D_1$. This set of vertices is clearly connected. Moreover, we claim that this set is a forcing set. To see why, observe that, initially, $A_1$ forces $B_1$, $C_1$ forces $D_k$, $D_1$ forces $D_2$, and $C_j$ forces $A_j$ for all $2\le j \le k$. After this sequence of forces, $D_j$ forces $D_{j+1}$ for $2\le j \le k-1$, since all of the $A_j$'s have been colored. Lastly, after $D_j$ is colored for $1\le j \le k$, $A_j$ will force $B_j$ for $2\le j \le k$. Hence, the given set is forcing as claimed, and has cardinality $k +2$, or equivalently $\frac{n}{4}+2$. See Figure \ref{fig4} for an illustration of this construction on the flower snark $J_{28}$.\qed

\subsection{Graphs with fixed forcing number}

We have already shown that for a given connected graph $G$, $F_c(G)=1$ and $F(G)=1$ if and only if $G$ is a path. It is also easy to see that for a given connected graph $G$, $F(G)=n-1$ if and only if $G=K_n$. Surprisingly, the same result does not hold for $F_c(G)$. The family of graphs $G$ for which $F_c(G)=n-1$ was recently characterized (\cite{CF Paper}, personal communication); we provide a shorter graph theoretic proof below. 

\begin{thm}
\label{t:upperbd}
Let $G=(V,E)$ be a connected graph of order $n\geq 2$. Then, $F_c(G)=n-1$ if and only if $G=K_n$, $n \ge 2$, or $G=K_{1,n}$, $n \ge 3$.
\end{thm}
\proof
If $G= K_n$ or $G= K_{1,n}$, it's easy to verify that $F_c(G)=n-1$. Now suppose $F_c(G)=n-1$. If $G$ has no separating set of vertices, then $G$ is a clique. We will show that if $G$ has a separating set $S\subset V$ then $G$ is a star. 

Let $S$ be a separating set of minimum cardinality, and suppose first that $|S|>1$. Let $V_1$ be the vertex set of a component of $G-S$; pick an edge $(s,w)$, $s\in S$, $w\in V_1$, and pick a vertex $v\in V-(S\cup V_1)$ such that $\{v,s\}$ is not a separating set. If $|S|>2$, it is clear that such a $v$ exists; if $|S|=2$, $G-s$ must be connected, so simply pick $v$ to be a non-articulation point of $G-s$ which is not in $S$ or $V_1$. Then, $V-\{v,s\}$ is a connected forcing set of $G$, since $N(w)\subset V_1\cup S$, so all of the neighbors of $w$ except $s$ are colored; thus, $w$ will force $s$ and then $v$ will get forced as well. Thus we have found a connected forcing set of size $n-2$, a contradiction.

Suppose now that $|S|=1$, and suppose first that there are multiple articulation points. Let $A$ and $B$ be two blocks of $G$ which are leaves of the block tree of $G$ and which are not incident to the same articulation point. Color every vertex except one non-articulation point in $A$ and one non-articulation point in $B$. This is clearly a connected forcing set -- a contradiction. Now suppose there is only one articulation point $p$. If any block $B$ incident to $p$ is not $K_2$, color everything except one (non-articulation) vertex in $B$ and one (non-articulation) vertex in a different block. Then, the uncolored vertex in $B$ will be forced by one of its neighbors in $B$, followed by the uncolored vertex in the other block -- a contradiction. Thus, every block incident to $p$ must be $K_2$, so $G$ is a star.
\qed


\section{Conclusion}

In this paper, we have introduced and studied the connected forcing number. This parameter is a sharp upper bound to several important graph parameters such as the forcing number, path cover number, and maximum nullity. Each of the latter parameters is NP-hard to compute, which motivates the following question:

\begin{quest}
\label{quest1}
Is computing the connected forcing number of a graph NP-hard?
\end{quest}

We expect the answer to Question \ref{quest1} to be affirmative, but if not, the connected forcing number can be used to efficiently approximate the aforementioned graph parameters.

Another open problem, related to Fact \ref{iff_fact}, is finding conditions on a graph $G$ which guarantee that some minimum forcing set of $G$ is connected, or guarantee that no minimum forcing set of $G$ is connected. In particular, in Proposition \ref{snark upper}, we have shown an upper bound to the connected forcing number of a flower snark. For small examples, it seems impossible to find a flower snark with forcing number less than this upper bound; this observation motivates the following question:

\begin{quest}
For a flower snark  $J_n$, is $F_c(J_n)=F(J_n)=\frac{n}{4}+2$? 
\end{quest}

Indeed, we may ask an even more general question of whether the connected forcing number of any snark equals its forcing number.

Finally, we have characterized graphs with extremal connected forcing numbers 1, and $n-1$. We believe that the techniques used in our proofs may be able to characterize other graphs with very large and very small connected forcing numbers. Thus, our last question is:

\begin{quest}
Which graphs satisfy $F_c(G)=2$, $F_c(G)=n-2$, $F_c(G)=3$, or $F_c(G)=n-3$?
\end{quest}

Note that the analogous problem of characterizing graphs with $F(G)=3$ and $F(G)=n-3$ is also open, while the problems of characterizing graphs with $F(G)=2$ and $F(G)=n-2$ have been solved in \cite{row} and \cite{AIM-Workshop} respectively, the latter using algebraic techniques.

\section*{Acknowledgement}
The first author's work is supported by the National Science Foundation under Grant No. 1450681.


\begin{thebibliography}{99}


\bibitem{AIM-Workshop}
AIM Special Work Group.
\newblock Zero forcing sets and the minimum rank of graphs.
\newblock {\em Linear Algebra and its Applications}, 428 (7): 1628--1648, 2008.


\bibitem{k-Forcing}
D. Amos, Y. Caro, R. Davila, and R. Pepper.
\newblock Upper bounds on the $k$-forcing number of a graph.
\newblock {\em Discrete Applied Mathematics}, 181: 1--10, 2015.


\bibitem{zf_tw}
F. Barioli,  W. Barrett, S.M. Fallat, T. Hall, L. Hogben, B. Shader, P. van den Driessche, and H. van der Holst.
\newblock Parameters Related to Tree-Width, Zero Forcing, and Maximum Nullity of a Graph.
\newblock {\em Journal of Graph Theory}, Volume 72 (2): 146--177, 2013.


\bibitem{Barioli}
F. Barioli, W. Barrett, S. Fallat, H. T. Hall, L. Hogben, B. Shader, P. van den Driessche, and 
H. van der Holst. Zero forcing parameters and minimum rank problems. \emph{Linear Algebra
and its Applications}, 433: 401--411, 2010.




\bibitem{benson}
K. Benson, D. Ferrero, M. Flagg, V. Furst, L. Hogben, V. Vasilevska, and B. Wissman.
\newblock Power domination and zero forcing.
arXiv:1510.02421, 2015.


\bibitem{bondy}
J. A. Bondy and U. S. R. Murty. \emph{Graph Theory with Applications}. Vol. 290. London, Macmillan, 1976.



\bibitem{quantum1}
D. Burgarth and V. Giovannetti.
\newblock Full control by locally induced relaxation.
\newblock{\em Physical Review Letters},  99 (10): 100501, 2007.


\bibitem{logic1}
D. Burgarth, V. Giovannetti,  L. Hogben,  S. Severini,  and M. Young.
\newblock Logic circuits from zero forcing.
\newblock {arXiv:1106.4403}, 2011.

\bibitem{butler}
S. Butler, L. DeLoss, J. Grout, H.T. Hall, T. McKay, J. Smith, and G. Tims. 
Minimum Rank Library (Sage programs for calculating bounds on the minimum rank of a graph, and for computing zero forcing parameters).


\bibitem{zf_np2}
C. Chekuri and N. Korula.
\newblock A graph reduction step preserving element-connectivity and applications.
\newblock {\em Automata, Languages and Programming}, 254--265. Springer 2009.

\bibitem{CF Paper}
R. Davila, M. Henning, C. Magnant, and R. Pepper.
\newblock Bounds on the connected forcing number of a graph.
\newblock In Preparation, 2016.

\bibitem{Edholm}
C. Edholm, L. Hogben, J. LaGrange, and D. Row.
\newblock Vertex and edge spread of zero forcing number, maximum nullity, and minimum rank of a graph.
\newblock {\em Linear Algebra and its Applications}, 436 (12): 4352--4372, 2012.


\bibitem{GJ}
Garey, M., D. Johnson, 
{\em Computers and Intractability}, 
W.H. Freeman \& Company, San Francisco, 1979.


\bibitem{Eroh}
L. Eroh, C. Kang,  and E. Yi.
\newblock Metric dimension and zero forcing number of two families of line graphs.
\newblock {\em arXiv:1207.6127}, 2012.




\bibitem{powerdom3}
 T. Haynes,  S. Hedetniemi, S. Hedetniemi, and M. Henning.
\newblock Domination in graphs applied to electric power networks.
\newblock {\em SIAM Journal on Discrete Mathematics}, 15 (4): 519--529, 2002.

\bibitem{Hogben}
L. Hogben. Minimum rank problems. \emph{Linear Algebra and its Applications}, 432: 1961--1974,
2010.

\bibitem{Huang}
L.-H. Huang, G. J. Chang, H.-G. Yeh. On minimum rank and zero forcing sets of a graph.
\emph{Linear Algebra and its Applications}, 432: 2961--2973, 2010.

\bibitem{row}
D. D. Row. 
A technique for computing the zero forcing number of a graph with a cut-vertex.
\emph{Linear Algebra and its Applications}, 436: 4423--4432, 2012.


\bibitem{Meyer}
S. Meyer.
\newblock Zero forcing sets and bipartite circulants.
\newblock {\em Linear Algebra and its Applications}, 436 (4): 888--900, 2012.


\bibitem{zf_np}
M. Trefois and J. C. Delvenne.
\newblock Zero forcing number, constrained matchings and strong structural controllability.
\newblock {\em  arXiv:1405.6222v2}, 2015.


\bibitem{powerdom2}
M. Zhao,  L. Kang, and G. Chang.
\newblock Power domination in graphs.
\newblock {\em Discrete Mathematics}, 306 (15): 1812--1816, 2006.



\end{thebibliography}
\end{document}